\documentclass[twocolumn,superscriptaddress,showpacs,floatfix,preprintnumbers]{revtex4-1}

\usepackage{array}
\usepackage{float}
\usepackage{lipsum}
\usepackage{natbib}
\usepackage[normalem]{ulem}
\usepackage[colorlinks,citecolor=blue,linkcolor=blue]{hyperref}
\usepackage{amsmath}
\usepackage{amssymb}
\usepackage{graphicx}
\usepackage{color}
\usepackage{epsfig}
\usepackage{amsmath}
\usepackage{amssymb}
\usepackage{epstopdf}
\usepackage{sidecap}
\usepackage{booktabs}
\usepackage{rotating}

\sidecaptionvpos{figure}{c}

\DeclareGraphicsExtensions{.png .jpg .pdf .eps}

\renewcommand{\i}{\mathrm{i}}
\newcommand{\ee}{\mathrm{e}}
\newcommand{\p}{\partial}
\newcommand{\mc}{\mathcal}
\newcommand{\mb}{\mathbf}
\newcommand{\vph}{\varphi}

\newcommand{\ex}{\hat{\mb{e}}_x}
\newcommand{\ey}{\hat{\mb{e}}_y}
\newcommand{\ez}{\hat{\mb{e}}_z}
\newcommand{\bs}{\boldsymbol}

\begin{document}

\title{Isolated and hybrid bilayer graphene quantum rings}

\author{M. Mirzakhani}\email{mirzakhani@ibs.re.kr}
\affiliation{School of Physics, University of the Witwatersrand, Johannesburg, Wits 2050, South Africa}
\affiliation{Center for Theoretical Physics of Complex Systems, Institute for Basic Science, Daejeon, 34126, South Korea}

\author{D. R. da Costa}\email{diego_rabelo@fisica.ufc.br}
\affiliation{Departamento de Fisica, Universidade Federal do
	Cear\'a, Campus do Pici, 60455-900 Fortaleza,
	Cear\'a, Brazil}

\author{F. M. Peeters}\email{francois.peeters@uantwerpen.be}
\affiliation{Department of Physics, University of Antwerp, Groenenborgerlaan 171, B-2020 Antwerp, Belgium}

\date{\today}

\begin{abstract}
	Using the continuum model, we
	investigate the electronic properties of two types of bilayer graphene (BLG)
	quantum ring (QR) geometries: (i) an isolated BLG QR and (ii) a monolayer graphene
	(MLG) with a QR put on top of an infinite graphene sheet (hybrid BLG QR).
	Solving the Dirac-Weyl equation in the presence of a perpendicular
	magnetic field and applying the infinite-mass boundary condition at the ring
	boundaries, we obtain analytical results for the energy levels and
	corresponding wave spinors for both structures.
	In the case of isolated BLG QR, we observe a sizeable and magnetically tunable 
	band gap which agrees with the tight-binding transport simulations.
	Our analytical results also show the intervalley symmetry 
	$ E^K_e (m) = -E^{K'}_h(m) $ between the electron (e)
	and hole (h) states ($ m $ being the angular momentum quantum number)
	for the energy spectrum of the isolated BLG QR.
	The presence of interface boundary in a hybrid BLG QR modifies drastically 
	the energy levels as compared to that of an isolated BLG QR. Its energy 
	levels are tunable from MLG dot, to isolated BLG QR, and to MLG 
	Landau energy levels as magnetic field is varied.
	Our predictions can be verified experimentally using different techniques 
	such as by magnetotransport measurements.
\end{abstract}

\maketitle


\section{Introduction} \label{intro}
Over the past years, a family of two-dimensional (2D) graphene  nanostructures,
including graphene nanoribbons
\cite{Son2006,Ezawa2006,Castro2008,Li2008,Sahu2008,Lam2009},
quantum rings (QRs) 
\cite{Recher2007,Russo2008,Huefner2009,Zarenia2009,daCosta2014R,Cabosart2017},
quantum dots (QDs) 
\cite{Ezawa2007,Bardarson2009,Wimmer2010,Giavaras2010,Rozhkov2011Re,Allen2012,
Zebrowski2013,Guclu2014Re,Bischoff2015Re,Bischoff2015,Mirzakhani2016ABC,
daCosta2015,daCosta2016dis,Mirzakhani2017el,Grujic2011},
and antidots \cite{Pedersen2008,Furst2009,Kim2010,Pedersen2012,Jessen2019} with
different type of geometries, edge types, and stackings of graphene layers
have received increasing interest.
These studies showed that the electronic and optical properties
of graphene QDs can be modified by size, shape, edge type,
and electrostatic gating; see, e.g., also Refs.\ \cite{Rozhkov2011Re,Guclu2014Re,
	Ezawa2007,Mirzakhani2017el,Giavaras2010}.
The effect of twisting on the electronic and transport properties of
bilayer graphene (BLG) nanostructures has also been recently addressed in
Refs.\ \cite{Landgraf2013,Suarez2014,Suarez2015,Pelc2015,Fleischmann2018,
	Tiutiunnyka2019,Mirzakhani2020}.
Within today's technology, such as nanolithography, 
it is possible to realize such 2D nanostructures on a scale of a few tens
\cite{Eroms2009,Ihn2010} or even only a few \cite{Ponomarenko2008,Xu2013} nanometers 
as well as, in different types of graphene layer stackings
\cite{Shih2011,Tsoukleri2015}.


Furthermore, bandstructure engineering can be performed by creating
periodic arrays of holes in both MLG
\cite{Jessen2019,Furst2009,Kim2010,Eroms2009}
and BLG \cite{Kvashnin2015}
sheets, known as graphene antidot lattices.
Depending on the size and the period of the holes, such graphene
nanostructures render graphene semiconducting with a sizeable
band gap which displays a wide range of electronic and optical
properties \cite{Kvashnin2015,Pedersen2008O,Gregersen2015}.
Very recently, a related system has been investigated where a MLG sheet
consists of highly regular triangularly arranged holes \cite{Jessen2019}.
The results showed that the structure displays both insulating behaviour
and ballistic transport in excellent
agreement with analytical calculations which describes the structure as
a quantum system consisting of connected \textquotedblleft Dirac
rings\textquotedblright\ \cite{Jessen2019,Thomsen2017}.
In this model, Dirac fermions are strongly confined in a MLG QR
geometry using an infinite-mass (IM) potential.
While in the case of MLG such holes can act as scattering centers,
several experimental works \cite{Liu2009,Zhan2011} demonstrated that
in the case of BLG, the adjacent
layers can connect with each other thus resulting in the formation of
periodic arrays of connected \textit{edgeless} Dirac rings in BLG structure.

\begin{figure}
	\centering
	\includegraphics[width = 8 cm]{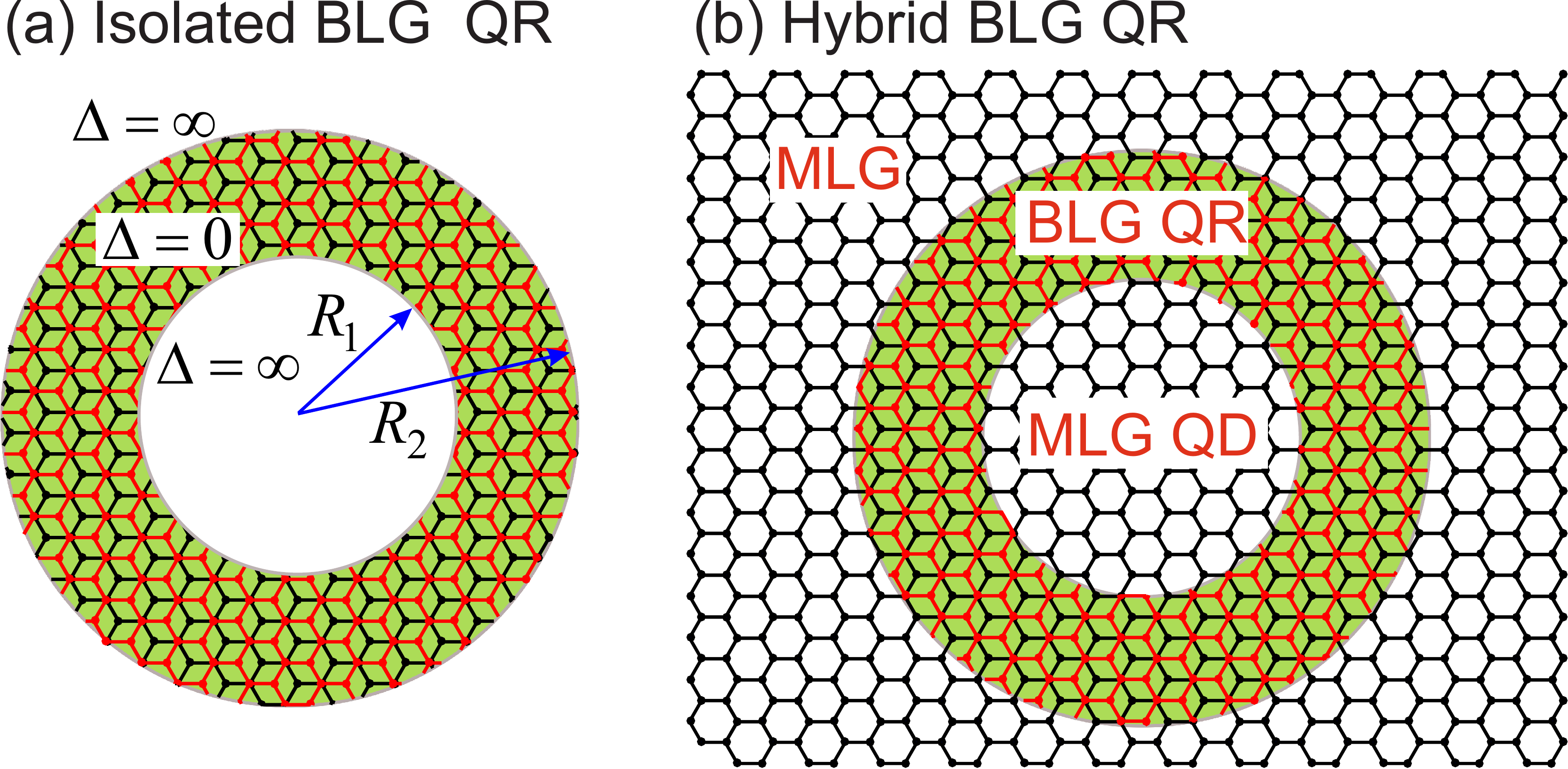}
	\caption{Schematic pictures of the proposed circular BLG QRs
		with inner and outer radii of $R_1$ and $ R_2 $, respectively.
		(a) Isolated BLG QR and (b) hybrid BLG QR sandwiched between MLG
		QD and MLG sheet.
	}
\label{fig1}
\end{figure}

Here, we aim to investigate the electronic properties of such rings in BLG
as an individual QR defined by an IM potential (isolated ring) as
shown in Fig.~\ref{fig1}(a).
Of course in graphene nanostructures, the type of edge plays an important role
and their effects are well known, in particular the existence of a zero-mode state
at zigzag edges \cite{Ezawa2007,Akola2008, Libisch2009}.
Using the IM boundary condition removes the edge effects and has the advantage 
that analytical results can be obtained while still representing a real 
system \cite{Grujic2011}.
Theoretically, the IM boundary condition for confining neutrinos
in a hard-wall billiard was derived by Berry and Modragon
\cite{Berry1987}.
This boundary condition was previously employed to investigate
the electronic properties of MLG nanostructures (dot, antidot, ring)
\cite{Grujic2011,Schnez2008,Pedersen2012,Thomsen2017},
BLG QDs \cite{daCosta2014D}, and trilayer graphene QDs
\cite{Mirzakhani2016ABC}.
Previous studies of BLG QRs based on Dirac equation includes
electrostatically defined BLG QRs \cite{Zarenia2009R,Zarenia2010R}
which was solved numerically in Ref.\ \cite{Zarenia2009R} and
was modeled as zero-width-ring geometry in Ref.\ \cite{Zarenia2010R}.
However, at present there is, to our knowledge, no theoretical
study on the energy spectrum of BLG QRs with IM boundary condition that
models a realistic ring.

In the present work, we solve the Dirac-Weyl equation in the presence of a
perpendicular magnetic field and apply IM boundary condition
at the ring boundaries to obtain analytical results for the energy
levels and corresponding wave functions.
In the case of isolated BLG QR, we find an excellent agreement between the 
analytical predictions for the size of the 
band gap as a function of the magnetic field with that found 
in the conductance for two-terminal QR device simulated using the KWANT 
package \cite{Kwant2014} 
based on Landauer-B\"uttiker formalism and the tight-binding model (TBM).
This is in contrast with previous study of zero-width BLG QR 
\cite{Zarenia2010R}, showing a fixed energy band gap as a function 
of magnetic field.
Our analytical results also show that the energy spectrum of the isolated
ring exhibits
intervalley symmetry $E^K_e(m) = -E^{K'}_h (m)$ for the electron
(e) and hole (h) states where $m$ is the angular momentum quantum number.

The agreement between the obtained analytical results and TB 
simulations prompted us to consider also a \textit{hybrid} 
BLG QR where a MLG ring is put on top of an infinite MLG sheet [Fig.~\ref{fig1}(b)].
Experimental realization of such 2D heterostructures can be
challenging but is doable within today's technology.  
Defining the ring layer by a staggered site-dependent IM potential 
(e.g., using an anti-ring shaped hexagonal boron nitride as a substrate) is
a way which can be used to realize the hybrid BLG QR. 
The hybrid structure can also be realized by (accidental) nanostructuring one 
of the graphene layers in BLG.
For instance, topographic images have revealed that multilayer samples
exfoliated from graphite often contain atomic steps and islands
of one or few layers of graphene \cite{Kobayashi2005,Niimi2006,
	Rutter2008,Clark2014,Yan2016}.
They have been previously investigated
both theoretically and experimentally, in
different configurations such as single MLG-BLG junction
\cite{Puls2009,Nakanishi2010,Koshino2010,Tian2013,Abdullah2016,Yin2017,
	Mirzakhani2018},
double MLG-BLG junctions (MLG-BLG-MLG) \cite{Yin2013,Mirzakhani2017MBM},
and hybrid QD structures \cite{Mirzakhani2016dju}.
In all these studies, the interface between MLG and BLG regions
was considered  as zigzag or armchair junctions which modifies considerably
the electronic properties of such structures.

A striking feature of the hybrid BLG QR is that the energy levels of the ring 
seem to \textit{interplay} between the MLG dot, isolated BLG QR, and MLG Landau energy levels 
as magnetic field increases.
In addition, as a function of the magnetic field, the energy spectrum of both 
structures exhibits Aharonov-Bohm (AB) oscillations.
We also investigate the dependence of the energy spectrum
on the ring width for both structures.


Finally, we analyze the valley- and layer-resolved local density of
states (LDOS)
for both proposed structures and our findings show that,
at a given magnetic field, the contributions of the valleys as well as that
of the layers in the LDOS can be different.
This feature can be used in valleytronics applications of such
graphene-based nanostructures if valley mixing is precluded.

\section{Theory and model} \label{theo}

We consider two different BLG nanostructures in the presence of
a perpendicular magnetic field: (i) isolated BLG QR defined by a site-dependent
staggered media [Fig.\ \ref{fig1}(a)] and
(ii) hybrid BLG QR sandwiched
between a MLG QD and an infinite MLG region, as shown in Fig.\ \ref{fig1}(b).
The latter one can also be regarded as an infinite MLG sheet on which a second
MLG ring is sitting on top of the first, thus realizing a BLG QR in the
AB-stacking (Bernal) configuration.
Dirac equation is solved for both MLG and BLG regions, with
appropriate boundary conditions.
By employing the IM boundary condition, we obtain analytical results for
the energy levels and corresponding wave functions in each structure.

Experimentally, such a mass potential can be induced
by sandwiching the BLG sheet between
substrates such that the $A$ and $B$ sublattices in each graphene
sheet feel different potentials \cite{Giovannetti2007,Bhowmick2011}.
Equivalently, graphene nanostructures that are etched out of graphene
sheets exhibit a strong confinement that can be modelled with
IM-boundary condition.

In the presence of a perpendicular magnetic field $\mb{B} = B \ez$,
the dynamics of carriers in the honeycomb lattice of carbon atoms of MLG
can be described by the following Hamiltonian  \cite{Recher2007},
\begin{equation}\label{eqn:dirac}
\mc{H} = v_{F}\bs{\Pi} \cdot \bs{\sigma} + \Delta(\mb{r}) \sigma_z,
\end{equation}
where $v_{F} \approx 10^6$ m/s is the Fermi velocity,
$\bs{\Pi} =  \mb{p} + e \mb{A}$ is the 2D kinetic momentum
operator  with $\mb{p} = -\i \hbar\, (\p_x,\p_y)$, $-e$ being the electron charge,
and $\mb{A} = (B/2)(-y\ex + x \ey)$ is the vector potential taken in the symmetric
gauge.
$\bs{\sigma} =(\sigma_x, \sigma_y, \sigma_z)$ denotes the Pauli matrices
and $\Delta(\mb{r})$ is a position-dependent mass term.
In polar coordinates $(r,\vph)$, the Hamiltonian \eqref{eqn:dirac}
reduces to the form
\begin{equation}\label{eqn:diracM}
 \mc{H} = E_0 \left(
 \begin{array}{cc}
 \delta & \Pi_{-} \\
 \Pi_{+} & -\delta \\
 \end{array}
 \right),
\end{equation}
 where $E_0 = \sqrt{2} \hbar\, v_F/ l_B$ is the cyclotron energy with
 $l_B = \sqrt{\hbar/ e B}$ the magnetic length,
 $ \delta = \Delta(\rho) / E_0$, and the momentum operator
 \begin{equation}\label{eqn:Pi}
    \Pi_{\pm} = \Pi_x \pm \i \Pi_y =
    -\i \ee^{\pm \i \tau \vph} \frac{1}{2}
    \left[ \frac{\p}{\p \rho} \pm \frac{\i \tau}{\rho} \frac{\p}{\p \vph} \mp \tau \rho \right].
 \end{equation}
 Here, $\rho = r / \sqrt{2}\, l_B$ is a dimensionless radial coordinate and
 $\tau = \pm 1$ distinguish the $K$ and $K'$ valleys.
 Because of circular symmetry, the two-component spinor wave function becomes
 $\Psi^{\tau}(\rho, \vph) =
 \ee^{\i m \vph} [\phi_A^\tau(\rho) ,\i \phi_B^\tau(\rho) \ee^{\i\! \tau \vph}]^T$,
 where the radial dependence of the spinor components is described by
\begin{subequations} \label{eqn:morad}
\begin{align}
 \frac{1}{2} \Big[\frac{\p}{\p\rho} + \frac{(\tau m + 1)}{\rho} + \tau \rho \Big] \phi_{B}^\tau(\rho) &= (\epsilon-\delta) \phi_{A}^\tau (\rho), \label{eqn:morada}\\
 \frac{1}{2} \Big[\frac{\p}{\p\rho} - \frac{\tau m}{\rho} - \tau \rho \Big]\phi_{A}^\tau (\rho) &= -(\epsilon + \delta)\phi_{B}^\tau (\rho) \label{eqn:moradb},
\end{align}
\end{subequations}
where $m = 0, \pm1,\pm2, \ldots$ denotes the angular momentum label and
$\epsilon = E/E_0$ being the dimensionless carrier energy.
Decoupling the above equations and using the ansatz for
$\phi_A^\tau(\rho) = \rho ^{-m} \ee^{-\rho^2/2} f (\rho^2)$,
one arrives at the associated Laguerre differential equation
($\tilde{\rho} = \rho^2$)
 \begin{equation}\label{eqn:AL}
    \tilde{\rho} f''(\tilde{\rho}) + (-m + 1 - \tilde{\rho}) f'(\tilde{\rho})
    + \lambda f(\tilde{\rho}) = 0,
 \end{equation}
where
 \begin{equation}
    \lambda = \frac{1}{2}[-(\tau+1) + 2(\epsilon^2 - \delta^2)].
 \end{equation}
The general solution to the associated Eq.\ \eqref{eqn:AL} is
\begin{equation}
   f(\rho^2) = C_1 L_\lambda^{-m}(\rho^2) + C_2 U(-\lambda, 1 - m,\rho^2),
\end{equation}
where the constants $C_1$ and $C_2$ are determined by the boundary conditions.
$U(a,b,x)$ is the \textit{confluent hypergeometric function of the second kind} and
$L_a^b(x)$ is the \textit{generalized Laguerre polynomial} which can be defined in
terms of the confluent hypergeometric function of the \textit{first} kind $M(a,b,x)$ [an alternatinig notation is $ _1 F_1(a,b,x) $] as
\begin{equation}
 L_a^b(x) = \left(\begin{array}{c} a+b \\ a
 \end{array}\right) M(-a,b+1,x),
\end{equation}
where $ \left(\begin{array}{c} a+b \\ a \end{array}\right) $ is the generalized binomial
coefficient.
Notice that $ \rho^{-m} \ee^{-\rho^2/2} L_\lambda^{-m}(\rho) $ and
$ \rho^{-m} \ee^{-\rho^2/2} U(-\lambda,1-m,\rho) $
converge to finite values in the limits $ \rho \rightarrow 0 $ and
$ \rho \rightarrow \infty $, respectively.
So, depending on the geometry of the graphene nanostructures (dot or antidot),
one can choose the appropriate wave functions to satisfy the corresponding
boundary conditions.

The other spinor component of the wave function, $\phi_B^\tau(\rho)$, can be obtained
using Eq.\ \eqref{eqn:moradb} and by employing the properties of
$U(a,b,x)$ and $L_a^b(x)$.
Thus the spinor components become
\begin{subequations}
\begin{multline}
  \phi_A^\tau(\rho) = \rho ^{-m} \ee^{-\rho^2/2}\ \times \\
    \Big[ C_1 L_\lambda^{-m}(\rho^2) + C_2 U(-\lambda, 1 - m,\rho^2)
    \Big],
\end{multline}
and
\begin{multline}
 \phi_B^\tau(\rho) = \frac{-\tau}{\epsilon + \delta}\rho ^{-m - \tau} \ee^{-\rho^2/2}
 \Big[C_1 (\lambda + \tau_+)^{\tau_+} L_{\lambda + \tau}^{-m - \tau}(\rho^2) \\
 - C_2(\lambda + \tau_+)^{\tau_-} U(-\lambda - \tau, -m + 1 - \tau,\rho^2)\Big],
\end{multline}
\end{subequations}
where we have defined $\tau_\pm = (1\pm \tau)/2$.

The BLG region can be described in terms of four sublattices, labeled $A1,B1$,
for the lower layer and $A2,B2$, for the upper layer.
We only include the coupling between two atoms stacked
on top of each other, e.g., $B1$ and $A2$, and ignore the small
contributions of the other interlayer couplings.
Additional terms only cause small effects such as trigonal wrapping and electron-hole
asymmetry on the energy levels 
\cite{McCann2006}.
The effective Hamiltonian is \cite{McCann2006,Nilsson2006}
\begin{equation}\label{eqn:HBi}
\mc{H}_B= E_0\left(
\begin{array}{cccc}
\delta & \Pi_{-} & 0 & 0 \\
\Pi_{+} & -\delta & \tilde{\gamma_1} & 0 \\
0 & \tilde{\gamma_1} & \delta & \Pi_{-} \\
0 & 0 & \Pi_{+} & -\delta \\
\end{array}
\right),
\end{equation}
where $\tilde{\gamma_1} = \gamma_1/E_0$, with $\gamma_1 \approx 0.4$ eV being
the nearest-neighbor interlayer coupling term.
Solving the Dirac equation $\mc{H}_B\Phi = E \Phi$ for the four-component
wave function
\begin{multline}
\Phi^{\tau}(r,\vph) = \ee^{\i m \vph} \times  \\
[\ee^{-\i \tau \vph} \phi_{A1}(\rho),\i \phi_{B1}(\rho),\i \phi_{A2}(\rho),
\ee^{\i \tau \vph} \phi_{B2}(\rho)]^T,
\end{multline}
the radial dependence of the spinor components in BLG are described by
\begin{subequations} \label{eqn:birad}
\begin{align} 
 & \frac{1}{2}\Big[\frac{d}{d \rho} + \frac{\tau m}{\rho} + \tau \rho \Big]
   \phi_{B1}^\tau(\rho) = (\epsilon - \delta) \phi_{A1}^\tau(\rho),  \\
 & \frac{1}{2}\Big[\frac{d}{d \rho} - \frac{\tau m - 1}{\rho} - \tau\rho \Big]
   \phi_{A1}^\tau(\rho) - \tilde{\gamma_1}\phi_{A2}^\tau(\rho) \notag \\
 & \hspace{4cm} = -(\epsilon + \delta) \phi_{B1}^\tau(\rho),  \\
 & \frac{1}{2} \Big[\frac{d}{d \rho} + \frac{\tau m + 1}{\rho} + \tau\rho \Big]
   \phi_{B2}^\tau(\rho) - \tilde{\gamma_1}\phi_{B1}^\tau(\rho) \notag \\
 & \hspace{4cm} = -(\epsilon - \delta) \phi_{A2}^\tau(\rho),  \\
 & \frac{1}{2} \Big[\frac{d}{d \rho} - \frac{\tau m}{\rho} - \tau \rho \Big]
   \phi_{A2}^\tau(\rho)  = (\epsilon + \delta) \phi_{B2}^\tau(\rho).
\end{align}
\end{subequations}
Decoupling the system of equations \eqref{eqn:birad} and using
$\phi_{A2}^\tau(\rho) = \rho ^{-m} \ee^{-\rho^2/2} g(\rho^2)$,
we arrive at the following associated Laguerre differential equation
($\tilde{\rho} = \rho^2$)
\begin{equation}\label{}
\tilde{\rho} g''(\tilde{\rho}) + (-m + 1 - \tilde{\rho}) g'(\tilde{\rho})
+ \alpha_\pm (\epsilon) g(\tilde{\rho}) = 0,
\end{equation}
where $ \alpha_\pm  (\epsilon) $ is given by
\begin{equation} \label{eqn:alpha}
\alpha_\pm (\epsilon) = \frac{1}{2}\left[ 2(\epsilon^2-\delta^2) -1
\pm \sqrt{1 + 4 \tilde{\gamma}^2 (\epsilon^2-\delta^2)} \right].
\end{equation}
Accordingly, similar to the MLG region, $\phi_{A2}^\tau(\rho)$ can be expressed in
terms of $L_a^b(x)$ and $U(a,b,x)$ as follows:
\begin{multline} \label{eqn:phA2}
\phi_{A2}^\tau(\rho) = \rho ^{-m} \ee^{-\rho^2/2} \ \times \\
\sum_{\mu=\pm} \Big[ C_1^\mu L_{\alpha_\mu}^{-m}(\rho^2)
+ C_2^\mu U(-\alpha_\mu, 1 - m,\rho^2)\Big],
\end{multline}
where the constants $C_1^\mu$ and $C_2^\mu$ are determined by the
boundary conditions.
The other spinor components of the wave function can be obtained using
Eqs.\ \eqref{eqn:birad} by inserting $\phi_{A2}^\tau (\rho)$ and employing the
properties of $U(a,b,x)$ and $L_a^b(x)$ functions.
It is possible to express the other components in a
compact form as follow:
\begin{widetext}
\begin{subequations}  \label{eqn:phother}
	\begin{align}
	& \phi_{A1}^\tau(\rho) = \frac{-\tau\, \rho ^{-m + \tau} \ee^{-\rho^2/2}}
	{\tilde{\gamma_1}(\epsilon^2 - \delta^2)}
	\sum_{\mu=\pm} \eta_\mu
	\Big[C_1^\mu (\alpha_\mu + \tau_-)^{\tau_-} L_{\alpha_\mu - \tau}^{-m + \tau}(\rho^2)
	- C_2^\mu (\alpha_\mu + \tau_-)^{\tau_+} U(-\alpha_\mu + \tau, -m + 1 +\tau,\rho^2) \Big],  \\
	& \phi_{B1}^\tau(\rho) = \frac{\rho ^{-m} \ee^{-\rho^2/2}}	{\tilde{\gamma_1}(\epsilon + \delta)}
	\sum_{\mu=\pm} \eta_\mu
	\Big[C_1^\mu L_{\alpha_\mu}^{-m}(\rho^2) + C_2^\mu U(-\alpha_\mu,-m + 1,\rho^2) \Big],   \\
	& \phi_{B2}^\tau(\rho)=\frac{\tau\, \rho ^{-m - \tau} \ee^{-\rho^2/2}}{\epsilon+\delta}
	\sum_{\mu=\pm}
	\Big[C_1^\mu (\alpha_\mu + \tau_+)^{\tau_+} L_{\alpha_\mu + \tau}^{-m - \tau}(\rho^2)
	- C_2^\mu (\alpha_\mu + \tau_+)^{\tau_-} U(-\alpha_\mu - \tau, -m + 1 -\tau,\rho^2) \Big],
	\end{align}
\end{subequations}
\end{widetext}
where
$ \eta_\mu = \big[\epsilon^2 - \delta^2 -(\alpha_\mu + \tau_+)\big] $.

\begin{figure*}
	\centering
	\includegraphics[width=17cm]{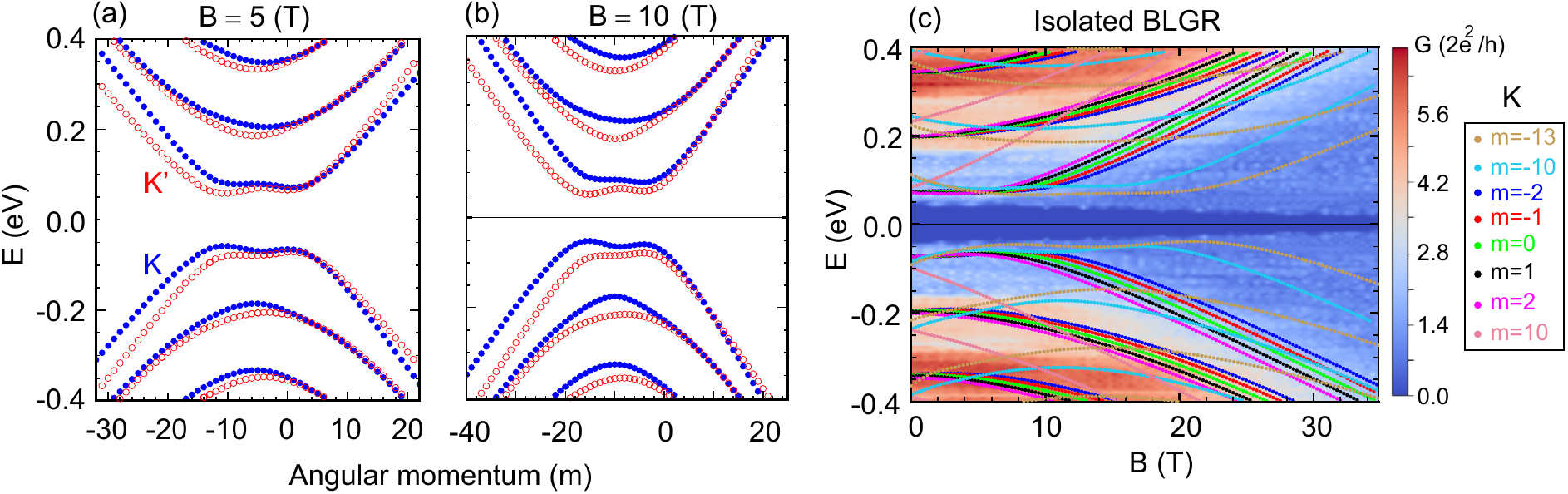}
	\caption[]{(a,b) Energy levels of an isolated BLG QR with inner and outer 
		radii $ R_1 = 30 $ nm and $ R_2 = 40 $ nm, respectively, as a 
		function of the angular momentum $ m $ for the $ K $ (solid blue circles)
		and $ K' $ (open red circles) valleys at the two different magnetic
		fields (a) $ B=5 $ T and (b) $ B=10 $ T.
		The symmetry $E^K_e(m) = -E^{K'}_h (m)$ is clearly visible.
		(c) $ K $-valley energy spectrum for the same ring as a function of the magnetic 
		field $ B $.
		The results are plotted for several angular momenta
		$ m =-13, \pm10, \pm2, \pm1, 0$.
		Background colored plot shows a 2D conductance of the same ring as 
		functions of $B$ and energy $E$ using KWANT package within the 
		Landauer-B\"uttiker formalism using the TBM.
	}
	\label{fig2}
\end{figure*}

Berry and Modragon derived the IM boundary condition for the
confinement of neutrinos in a hard-wall billiard described by the Dirac-Weyl
equation \cite{Berry1987}.
Lets consider a particle restricted in the plane $ \mb r = (x,y) $
subjected to a mass term potential
$ \Delta(\mb r) $, which is vanishing inside a certain domain and equal
to $ \Delta \rightarrow \infty $ outside it.
Solving the Dirac equation, Eq.\ \eqref{eqn:dirac}, for a two-component spinor
$ [\psi_1(\mb r),\psi_2(\mb r)]^T $ leads to the following relation at the
domain edge \cite{Berry1987,Schnez2008}
\begin{equation} \label{eqn:im}
\psi_2(\mb r)/\psi_1(\mb r) = \i \ee^{\i \tau \theta},
\end{equation}
where $ \theta $ is the polar angle of the normal vector pointing outward
from the domain boundary.
Within the next section, we will calculate the energy spectrum of both ring structures using the above-mentioned boundary condition.

It is worth mentioning that the IM boundary condition [Eq.~\eqref{eqn:im}]
does not necessarily imply that 
the total wave function, as well as the wave function components, go to zero at 
the boundaries of the quantum confined system. 
The use of similar Berry and Mondragon-like boundary condition to explain 
experimental measurements in graphene-based quantum confinement nanostructures has 
been reported in the literature 
\cite{Recher2007,Jessen2019,Ponomarenko2008,Schnez2008}, 
showing a good agreement between 
the predicted theoretical results and the electronic
properties measured experimentally. 
Furthermore, BLG based quantum systems with reconstructed edges present 
different types of edge, in addition to the conventional armchair and zigzag
terminations, leading to different boundary conditions different from that assumed 
here \cite{Akhmerov2008,Koskinen2008,Koskinen2009,Rakyta2010,Ostaay2011}.


\section{Numerical results} \label{Nr}

\subsection{Analytical calculations} \label{dirac}

In the case of isolated BLG QR, applying the IM boundary condition \eqref{eqn:im}
for each layer,
the spinor components at the radial distances
$R_1$ ($ \theta = \pi + \varphi $) and $R_2$ ($ \theta =\varphi $)
satisfy the conditions
\begin{subequations} \label{eqn:imbiR1}
\begin{align}
 & \phi_{B1}^\tau(\rho_1) + \phi_{A1}^\tau(\rho_1) = 0\,, \\
 & \phi_{B2}^\tau(\rho_1) - \phi_{A2}^\tau(\rho_1) = 0,
\end{align}
\end{subequations}
and
\begin{subequations} \label{eqn:imbiR2}
\begin{align} 
 & \phi_{B1}^\tau(\rho_2) - \phi_{A1}^\tau(\rho_2) = 0\,, \\
 & \phi_{B2}^\tau(\rho_2) + \phi_{A2}^\tau(\rho_2) = 0\,,
\end{align}
\end{subequations}
respectively, with $ \rho_i = R_i/ \sqrt2\, l_B $ ($ i=1,2 $).
The eigenvalue condition is determined by inserting the obtained spinors
for BLG, Eqs.\ \eqref{eqn:phA2} and \eqref{eqn:phother}, with $ \delta = 0 $
into the above four equations.
Note that the effect of IM potential $ \Delta \rightarrow \infty $ is
now expressed by the boundary conditions \eqref{eqn:imbiR1} and \eqref{eqn:imbiR2}.

For the hybrid BLG QR, the spinors corresponding to the sublattices in
the lower layer are continuous at the ring boundaries,
while the spinor components of the upper (ring) layer, $ \phi_{A2}^\tau (\rho)$
and $ \phi_{B2}^\tau (\rho) $, satisfy the IM boundary condition expressed
by Eqs.\ \eqref{eqn:imbiR1} and \eqref{eqn:imbiR2}.
Thus, in this case, the boundary conditions at $R_{1(2)}$ read
\begin{subequations}\label{eqn:imhyR1}
\begin{align} 
& \phi_{A}^\tau(\rho_{1(2)}) - \phi_{A1}^\tau(\rho_{1(2)}) = 0\,,  \\
& \phi_{B}^\tau(\rho_{1(2)}) - \phi_{B1}^\tau(\rho_{1(2)}) = 0\,,  \\
& \phi_{B2}^\tau(\rho_{1(2)}) \mp \phi_{A2}^\tau(\rho_{1(2)}) = 0,
\end{align}
\end{subequations}
where $ -(+) $ is used at $ R_{1(2)} $ boundary.

Here, we have to stress that,
using the above boundary conditions when implementing them
numerically gives two sets
of energy levels.
A set of levels corresponds to the pristine BLG Landau levels (LLs),
which mathematically originates from the proportionality of $L_a^b(x)$
and $U(a,b,x)$ functions when $ a $ ($ \equiv \alpha_\pm  (\epsilon) $)
becomes an integer number $ n $.
In this case, $L_n^b(x)= \frac {(-1)^n}{n!} U(-n,b+1,x)$ and the wave spinors
become finite in both limits $ r \rightarrow 0 $ and $ \infty $ as for
bulk BLG LLs \cite{Koshino2010}.
\begin{figure*}
	\centering
	\includegraphics[width=17cm]{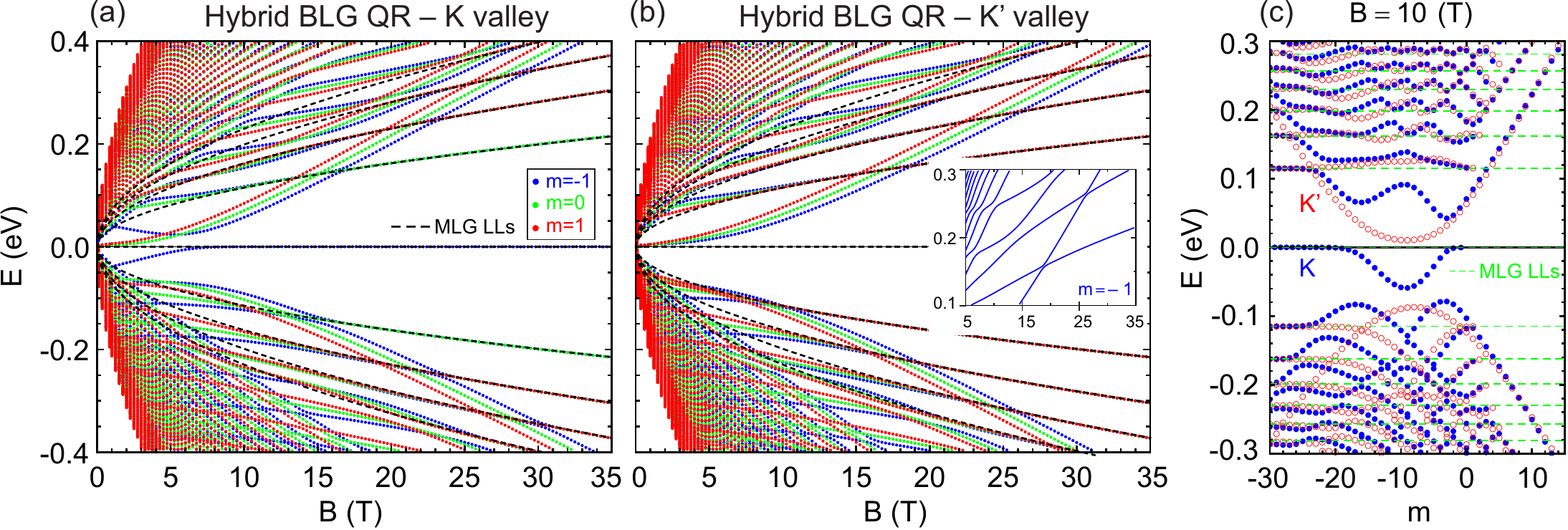}
	\caption[qw]{Energy spectrum of a hybrid BLG QR as a
		function of magnetic field $ B $, with inner and outer radii
		$ R_1 = 30 $ nm and $ R_2 = 40 $ nm, respectively, for the 
		(a) $ K $ and (b) $ K' $ valleys.
		The results are plotted for three different angular momenta
		$ m = \pm1, 0 $.
		The dashed black curves show the Landau levels ($ n=0,1, \ldots 4 $) of
		pristine MLG sheet.
		The inset in panel (b) shows an enlarged view of the
		spectrum, showing the anticrossings for angular momentum $ m = -1 $
		and the $ K' $ valley.
		(c) Energy levels as a function of the angular momentum $ m $ for the same
		hybrid ring at $ B=10 $ T for the two valleys
		$ K $ (solid blue circles) and $ K' $ (open red circles).
		The dashed green lines pertain to the bulk MLG LLs at $ B=10 $ T for
		$ n=0,1, \ldots 6 $.
	}
	\label{fig3}
\end{figure*}

In Figs.\ \ref{fig2}(a) and \ref{fig2}(b),  we plot the energy spectrum of both valleys
as a function of the angular momentum $ m $.
The results are presented for two different magnetic fields
(a) $ B=5 $ T and (b) $ B=10 $ T with solid blue (open red) circles for the
$ K $ ($ K' $) valley.
The inner and outer ring radii, respectively, are $ R_1=30 $ nm and $ R_2=40 $ nm.
Irrespective of the magnetic-field strength, the energy spectra exhibit intervalley
symmetry $E^K_e(m) = -E^{K'}_h (m)$ between the electron and hole states,
indicating that valley degeneracy is lifted.
Lifting the valley degeneracy, due to the presence of the magnetic field,
has also been noticed before in other graphene nanostructures
\cite{Recher2007,Mirzakhani2016ABC,daCosta2014D}, and is of great
interest because it could make them promising candidates for \textit{valleytronics}
applications.
One can see that decreasing the field $ B $, suppresses
the above-mentioned intervalley symmetry and the valley degeneracy is restored
at $ B=0 $ [cf. Figs. \ref{fig2}(a) and \ref{fig2}(b)].
In the case of zero-width BLG QR \cite{Zarenia2010R},
the energy spectrum shows only two energy levels for each $ m $,
and the spectrum versus $ m $ is very similar to the band structure of
a biased BLG sheet (see Fig.\ 10 in Ref.\ \cite{Zarenia2010R}).
However, here, we find several energy levels as a function of $ m $ and
the number depends on the width of the ring.

The energy spectrum of an isolated BLG QR, as a function of magnetic field,
is shown in Fig.\ \ref{fig2}(c) for several values of the angular momenta,
$m = -13$, $m = \pm 10$, $m = \pm 2$, $m = \pm 1$, and $m = 0$ at the $K$ valley
for the same ring parameters as in Figs.\ \ref{fig2}(a) and \ref{fig2}(b).
For a single valley, as seen in panel \ref{fig2}(a), two broken symmetries are
clearly visible in
the presence of the magnetic field: (i) For a specific angular momentum $ m $, the
electron-hole (e-h) symmetry is broken, i.e., $ E_e(m) \neq -E_h(m) $
and (ii)  $ E_i(m) \neq E_i(-m) $ ($i = e, h$), since time-reversal symmetry 
(TRS) is broken by the magnetic field \cite{Recher2007}.
Both symmetries are restored at $ B=0 $ as expected.
For large field strength, the magnetic-field dependence of the spectrum
becomes approximately linear which was also observed for the MLG ring and
antidot spectra \cite{Thomsen2017}.
Note that, as function of the magnetic field,
the lowest electron and hole energy levels show a
\textquotedblleft Mexican-hat\textquotedblright\ shape
similar as the Mexican-hat-shaped low-energy dispersion for a biased
AB-stacked BLG \cite{McCann2006R}, which is due to the fourth-order character
of the dispersion relation [Eq.~\eqref{eqn:alpha}].
Including the energy levels of both valleys [see Fig.\ \ref{fig6}(b)]
we can see that the confinement-induced band gap closes with increasing
magnetic field.
This is in contrast with the previous study of the zero-width BLG QR
\cite{Zarenia2010R},
which shows a constant energy band gap as a function of the magnetic field.
A similar result for the strong dependence of the energy gap on the magnetic field
in the finite-width MLG QR as well as a comparative study with the ideal 
zero-width one was also discussed in Ref.\ \cite{daCosta2014R}.

To see the connection between the obtained analytical energy levels and 
observable physical quantity like conductance, we show
in Fig.\ \ref{fig2}(c), a 2D color plot of two-terminal conductance $G$ of the same 
ring as functions of $B$ field and energy  using the KWANT package \cite{Kwant2014} 
within the Landauer-B\"uttiker formalism using the TBM 
(for details of the the TB simulation see Sec.\ \ref{tbm}).
We find excellent agreement between the analytical results and the 
tight-binding transport simulations.
Specifically, the closing of the band gap as a function of $B$ field in 
both calculations is seen to agree remarkably well. 
In addition, one can see high conductance values for large energies
($ |E| \gtrsim 0.2 $ eV) indicating the presence of a large number of conducting 
channels at these energies.
This is consistent with the analytical results where we find a high concentration
of energy states at high energies.
Notice that, in Fig.\ \ref{fig2}(c), we plotted only the energy levels 
for a few values of $m$ and only for one $K$ valley [a complete spectrum is shown
in Fig.\ \ref{fig6}(b)].

\begin{figure*}
	\centering
	\includegraphics[width=18cm]{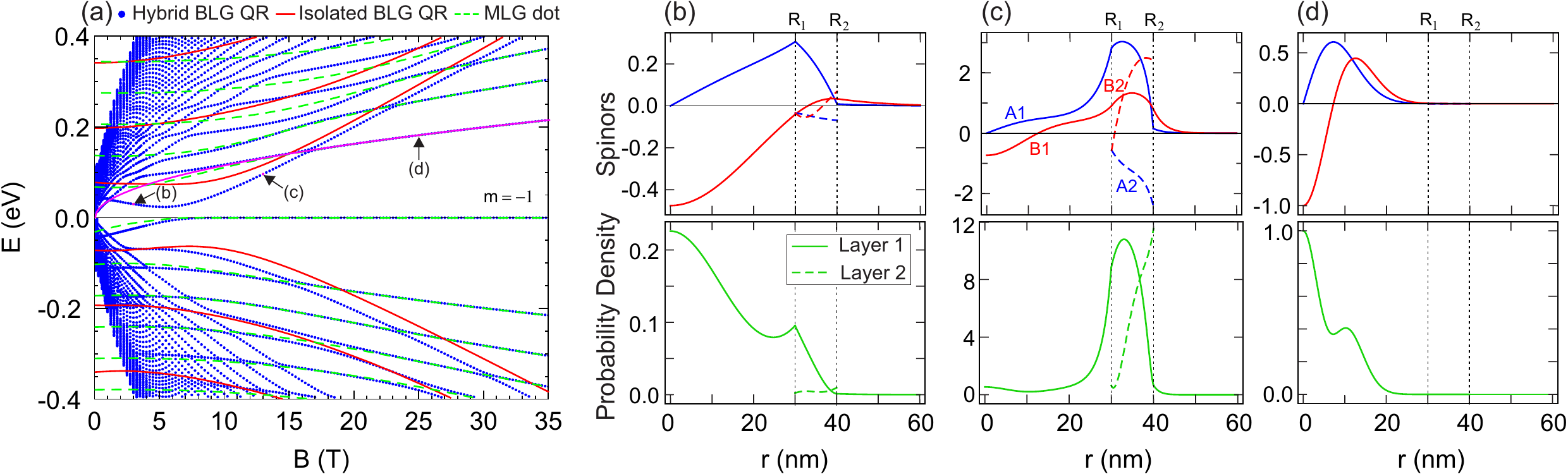}
	\caption[qw]{(a) Lowest energy levels of a hybrid BLG QR (blue circles),
		isolated BLG QR (solid red curves), and MLG dot (dashed green curves)
		as a function of the magnetic field $ B $ for
		angular momentum $ m=-1 $ in the $ K $ valley.
		The rings radii are $ R_1 = 30 $ nm and $ R_2 = 40 $ nm and the
		radius of the MLG dot is $ R_1 = 30 $ nm.
		The magenta curve indicates the first electron LL ($ n=1 $) of MLG sheet.
		Panels (b), (c), and (d) are the wave spinors (upper panels) and probability densities (lower panels) corresponding to, respectively, the states labeled
		by (b), (c), and (d) in the energy spectrum of panel (a).
		Layer 1 (2) is represented by the solid (dashed) curves.
		Sublattices $ Ai $, ($ Bi $) are represented by blue (red) curves.
	}
	\label{fig4}
\end{figure*}

Results for the energy spectrum of a hybrid BLG QR, of radius
$ R_1=30 $ nm and $ R_2=40 $ nm, as a function of magnetic field $ B $
at both valleys $ K $ and $ K' $ are shown in Figs.\ \ref{fig3}(a) and
\ref{fig3}(b), respectively.
The angular momenta are $ m = -1 $ (blue), $ m = 0 $ (green), and $ m = 1 $ (red).
The $ B $-field dependence of the hybrid BLG QR energy spectrum is
strikingly different from that of the isolated BLG QR.
Here, for $ B = 0 $, the spectrum is continuous because of the presence of
the infinite MLG, and with increasing magnetic field the degeneracy of
the states is lifted.
The discrete levels (reflecting the confined states)
depend on the angular momentum and the $ B $-field strength.
The spectra for both valleys show anticrossings, which are due to the
influence of the MLG dot and BLG QR interface [see the inset of Fig.\ \ref{fig3}(b)].
As a result of this interface, the symmetry condition $E^K_e(m) = -E^{K'}_h (m)$
(which holds for an isolated ring) is no longer preserved here, as shown
explicitly in Fig.\ \ref{fig3}(c).
At high magnetic fields, i.e., $ l_B < R_1 $, the energy levels merge into
the LLs of pristine MLG [$ E_n = \pm\sqrt{2n}\, \hbar v_F/ l_B $ with
$ n=0,1,\ldots $, shown by dashed black curves in Figs.\ \ref{fig3}(a) and \ref{fig3}(b)]
indicating that magnetic confinement dominates and the carriers become
localized at the center of the dot region (see Fig.\ \ref{fig4} and the
corresponding discussion).
It is also worth noting that, here, the states with $m < 0$ contributes to
the zero LL ($ n=0 $) only in one valley ($ K $).
This is in contrast with the other graphene nanostructures such as
MLG \cite{Grujic2011} and BLG QDs \cite{daCosta2014D} with
IM boundary condition in which the states from both valleys form the zero LL.
Notice that, unlike to the isolated BLG QR, it is not straightforward to calculate
the conductance of the hybrid QR in the TBM, since its bottom layer consists of an infinite graphene layer sheet.
However, in Sec.\ \ref{tbm}, we will compare explicitly the analytical energy levels with 
those obtained within the TBM for both ring structures.

\begin{figure}
	\centering
	\includegraphics[width=8.5cm]{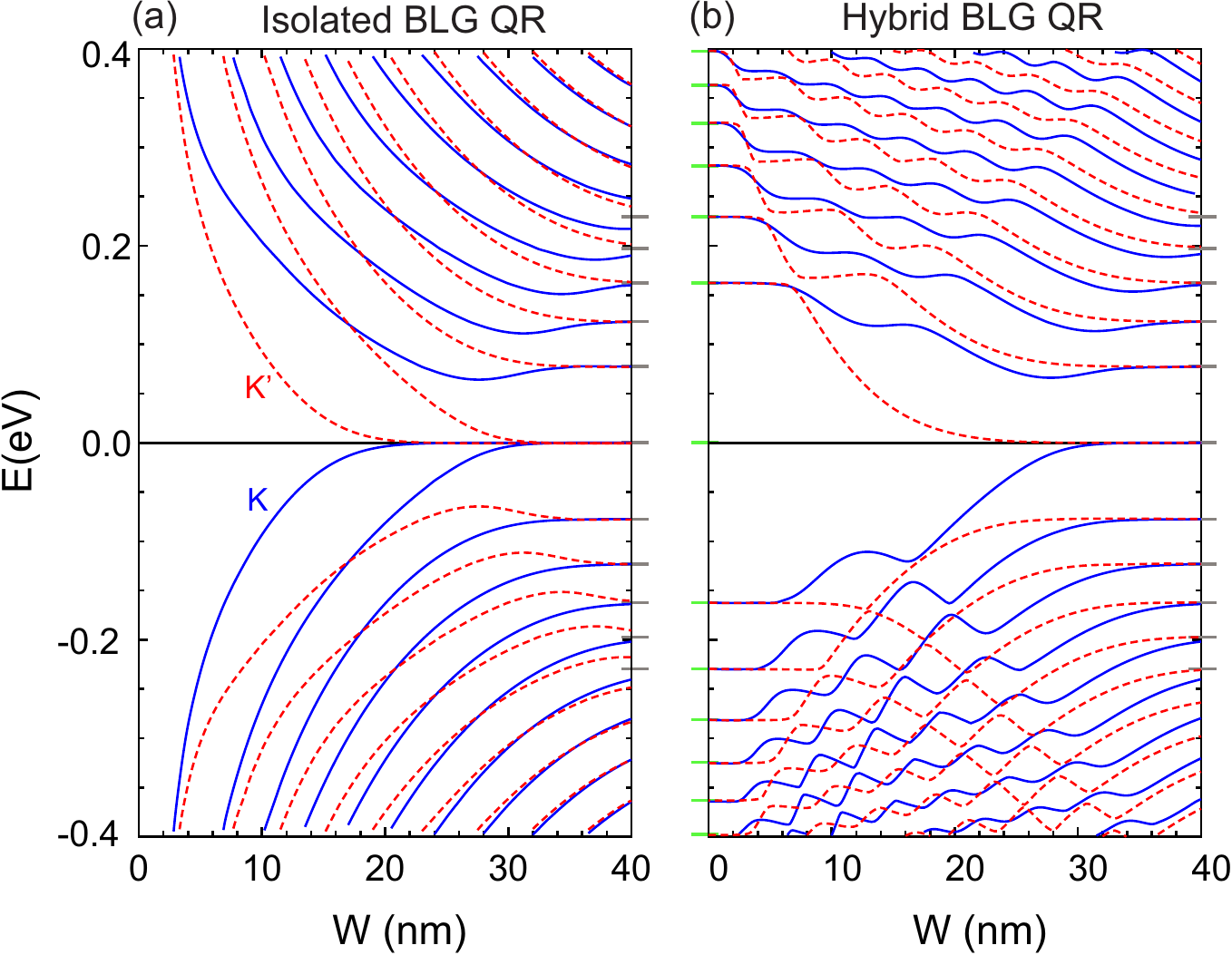}
	\caption[qw]{Energy levels of (a) isolated and (b) hybrid rings as a
		function of ring width $ W $, for angular momentum
		$ m=-20 $ at the two valleys $ K $ (solid) and $ K' $ (dashed)
		when $ B = 20 $ T.
		In both cases, the inner radius of the ring is $ R_1 = 20 $ nm
		and the outer one $ R_2 = R_1 + W $.
		The (horizental) solid green and gray lines indicate, respectively, the bulk
		MLG and BLG LLs at $ B=20 $ T.
	}
	\label{fig5}
\end{figure}

Figure \ref{fig3}(c) shows the energy spectrum as a function of the angular momentum
at a specific magnetic field 
($ B = 10 $ T) for both valleys $ K $ (solid blue circles) and $ K' $ 
(open red circles).
As seen, there is no symmetry between the e-h energies as well as between the valleys.
For a given magnetic field, the energy levels at both valleys are affected by
the ring interface for a particular range of $ m $ and converge
to the MLG LLs for larger $ m $'s (green lines), indicating carrier localization
inside the dot region.
Note that at the $ K $ ($ K' $) valley, the maximum value of the angular
momentum $ m_{\mathrm{max}} $ that is converged to the $ n $th electron LL of
MLG is $ m_\mathrm{max} = n-1 $ ($ m_\mathrm{max} = n,\ n \neq 0 $).
Further, in both valleys, the electron states show a smooth oscillatory
behavior as function of magnetic field
and strong anticrossings are visible also for hole states.
Similar behavior was also noted for the double MLG-BLG junctions studied in
Ref.\ \cite{Mirzakhani2017MBM} when the spectra are plotted as a function of the center coordinate of the cyclotron orbit.

\begin{figure*}
	\centering
	\includegraphics[width=17cm]{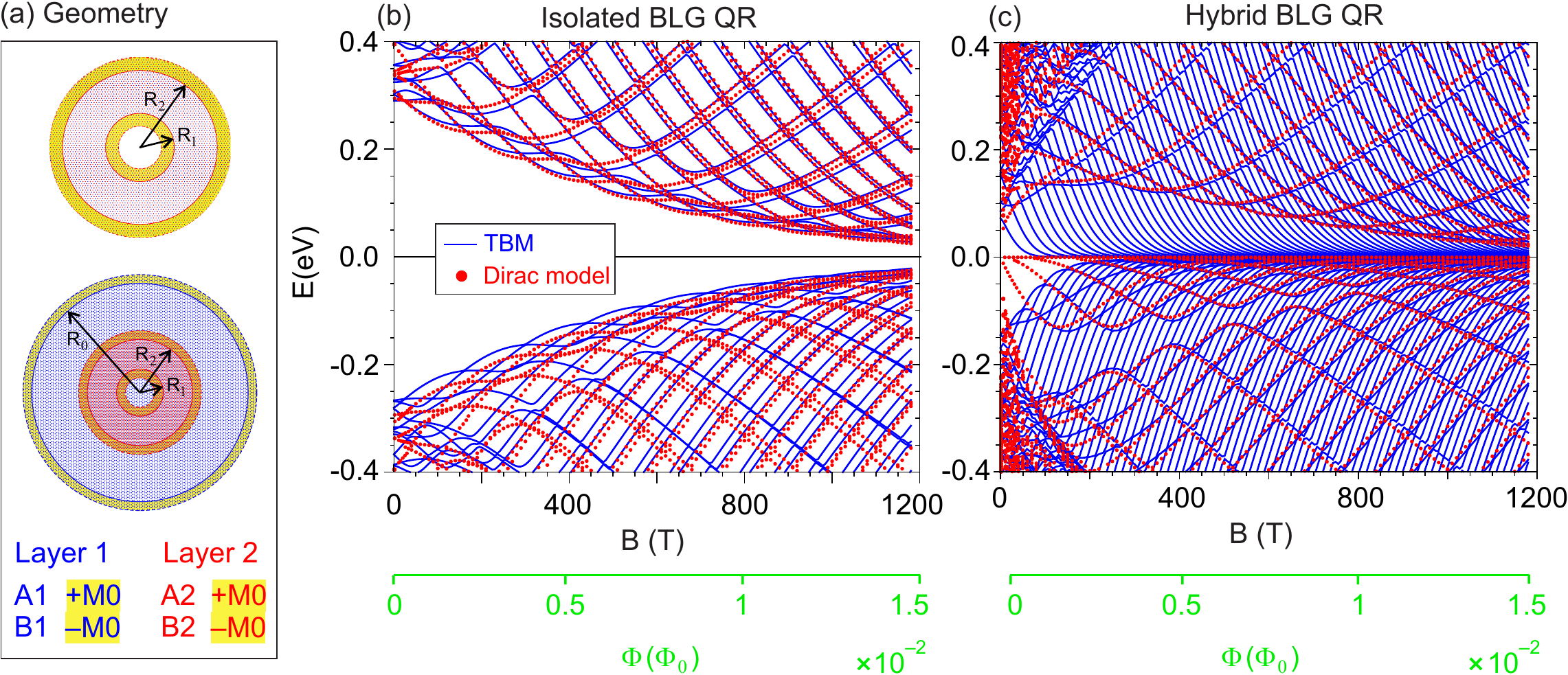}
	\caption[qw]{(a) Schematic geometry of the circular BLG QRs with inner and
		outer radii of $ R_1 $ and $ R_2 $, respectively, defined by the IM boundary
		used in TBM calculation.
		We define the honeycomb lattice of BLG with the lattice vectors 
		$a_1 = a_0 (\sqrt{3}/2, 1/2)$ and $a_2 = a_0 (\sqrt{3}/2, -1/2)$, 
		where $a_0 = 0.246$ nm is the lattice constant.
		The coordinates $x$ and $y$ are defined along the armchair and zigzag 
		directions, respectively (Fig.\ \ref{fig1}). 
		The upper (lower) panel depicts the isolated (hybrid) BLG QR geometry.
		The atoms of the two layers are represented by blue (layer 1) and
		red (layer 2) circles.
		The ring regions are surrounded by a site-dependent staggered potential
		(yellow	areas), where the atoms belonging to the sublattices
		$ A1 $ ($ A2 $) and	$ B1 $ ($ B2 $) have mass-term potentials
		of $ +M_0 $ ($ +M_0 $) and $ -M_0 $ ($ -M_0 $),
		respectively.
		In the case of hybrid ring, to simulate the lower layer as an infinite
		graphene sheet, we consider it as a large circular flake (layer 1)
		on which a second MLG ring (layer 2) is sitting on top of it.
		Further, to eliminate the specific edge effects of layer 1, here also,
		we apply the staggered potential on the atoms located within the
		ribbon width of  $ \Delta r = 2 a_0 $ 
		at the edge of the flake (yellow region).
		(b,c) Energy levels of (b) an isolated BLG QR and (c) a hybrid BLG QR
		as a function of the
		magnetic field $ B $, calculated within the Dirac (red circles)
		and the TB (solid blue curves) models.
		For both cases, the results are presented for rings with inner
		and	outer radii of $ R_1 = 8\, a_0 = 1.97 $ nm and
		$ R_2 = 18 a_0 = 4.43 $ nm.
		The mass potential is $ M_0 = 2 $ eV.
		In the hybrid ring, we use a circular flake of radius
		$ R_0 = 38 a_0 = 9.35 $ nm for the lower layer.
	}
	\label{fig6}
\end{figure*}

To better understand the behavior of the energy levels in the spectrum of
the hybrid ring, we plot in Fig.~\ref{fig4} the energy levels
of both types of rings and MLG dot for a specific angular momentum,
$ m = -1 $ [panel \ref{fig4}(a)],
as well as the corresponding spinor components [$ \phi_{\nu} (\rho) ,\ \nu=A,B, \ldots $]
and probability densities
[$ \phi_{A}^2 (\rho) + \phi_{B}^2 (\rho),\ldots $] for the representative energy
states marked by (b), (c), and (d) in Fig.\ \ref{fig4}(a), [panels (b)-(d)].
As seen, the energy levels of the hybrid ring seem to be tunable from
the MLG dot, to isolated BLG QR, and to MLG Landau energy levels as $ B $ increases.
At the representative state (b), the energy level (blue circles)
resembles that of the MLG dot (dashed green curves),
the corresponding quantum state is mostly confined inside the dot region with
a substantial probability density in the ring region, see \ref{fig4}(b) panels.
When the energy level approaches the one of the isolated ring
(red curves), e.g., point (c), the carrier is mostly confined inside the
ring region [Fig.\ \ref{fig4}(c)].
For strong magnetic field, e.g., point (d), the energy level
converges to the MLG LL (magenta curve), the state is completely confined 
inside the dot region, as seen in Fig.\ \ref{fig4}(d).

It is also interesting to investigate the effect of ring width on the energy
spectrum of both structures.
For this purpose, we plot in Fig.\ \ref{fig5}, the lowest energy levels as a
function of the width of the ring $ W $ ($ =R_2-R_1 $) for (a) isolated and
(b) hybrid BLG QRs.
In both cases, for the sake of clarity, the results are presented only for
particular angular momentum $ m=-20 $ at $ B=20 $ T.
In the case of an isolated ring, the energy levels are
separated by a gap, depending on the angular momentum and ring width,
which closes when $ W $ increases and the energy levels for both valleys approach
the LLs of bulk BLG as depicted by the horizontal gray lines.
In the hybrid ring, Fig.\ \ref{fig5}(b), at small ring widths, the
energy levels correspond to the LLs of MLG.
When $ W $ increases, the electron (hole) energy levels, due to the influence
of MLG-BLG interface, exhibit a flat plateau (strong anticrossing) features
for certain ranges of $ W $, and eventually merge into the LLs of bulk BLG.
The plateau-like and oscillatory features appearing in the energy spectrum
can be understood as hybridization of the energy levels of the terminated systems,
MLG QD and BLG antidot.
Similar behavior was found for previous studies of MLG-BLG junctions
\cite{Koshino2010,Mirzakhani2017MBM,Mirzakhani2016dju, daCosta2016dis}.

\subsection{Comparison with tight-binding model} \label{tbm}

In order to check the validity of the continuum approximation, we compare 
explicitly our
analytical results with the energy levels calculated within a nearest-neighbor 
TB approach.
We include only the nearest-neighbor hopping parameters
$ \gamma_0 = -2.7 $ eV and $ \gamma_1 = 0.4 $ eV as \textit{intralayer}
and \textit{interlayer} couplings, respectively.
The effect of an external magnetic field can be introduced into the calculations
via the Peierls substitution
$t_{ij} \rightarrow t_{ij} \ee^{\i 2\pi \Phi_{ij}}$,
where
$ \Phi_{ij}= (1/\Phi_0) \int^{\mb{R}_j}_{\mb{R}_i}
 \mb{A}(\mb{r})\cdot d\mb{r}$
is the Peierls phase \cite{Peierls1933} with
$\Phi_0 = h/e \approx 4.14 \times 10^{-15}$ Wb
the magnetic flux quantum.
The vector potential
corresponding to the external magnetic field $\mathbf{B} = B \mathbf{\hat{z}}$ perpendicular to the BLG flakes is chosen in the Landau gauge
$\mb{A}(\mb{r}) = (0, Bx, 0)$ for which one finds that $\Phi_{ij}$ is only nonzero
in the $y$ direction and is given by 
$\Phi_{ij} =\text{sign}(y_j - y_i) \frac{x_j + x_i}{2\sqrt{3} a_0} \frac{\Phi}{\Phi_0} $,
where $\Phi = B(\sqrt{3} a_0^2/2)$ is the magnetic flux threading one carbon 
hexagon ($a_0$ is the graphene lattice constant).
Here, the QRs are defined by a staggered site-dependent
potential such that the atoms belonging to the sublattices
$ A1 $ ($ A2 $) and $ B1 $ ($ B2 $) have a mass-term potential
of $ +M_0 $ ($ +M_0 $) and $ -M_0 $ ($ -M_0 $), respectively, as shown in
Fig.\ \ref{fig6}(a).
This simulates the substrate effect and can be regarded as the IM boundary
condition in the TBM.
More details of the geometries are provided in the caption of Fig.\ \ref{fig6}.
Furthermore, our calculations show that in the Dirac model, adopting the same
radius used to define the atomic geometries in TBM, results in the energy levels
which are slightly larger (i.e., shifted up) than those obtained by TBM.
Two possible explanations for this discrepancy can be considered.
First, in the TBM to simulate the IM boundary, we have used a narrow ribbon
of $ \Delta r $ containing atoms with different \textit{finite} on-site potentials 
in addition to the ring radii as shown in Fig.\ \ref{fig6}(a).
The second which has also been proposed in Ref.\ \cite{Pedersen2017}, can be attributed
to the $ \pi $ electrons that in the TBM extend over the whole geometry.
Accordingly, to compensate for this difference, we use a larger width
for the rings in the Dirac model.
Also note that, due to the computational limitation of the TBM, 
we have considered the small sizes of the rings and the applied magnetic 
field in our calculations is too large to be achievable in experiments.
However, for the study of the electronic properties of graphene nanostructures 
in the presence of a perpendicular magnetic field, one can define a scaling 
factor and thus extend the results to lower magnetic field and larger sample 
sizes, e.g., see Refs.\ 
\cite{Ferry2010,Cabosart2017,Thomsen2017}. 

In Figs.\ \ref{fig6}(b) and \ref{fig6}(c), we compare the results obtained
within the TB and Dirac
approaches for the lowest-energy levels of both rings as a function of
magnetic field $ B $, see the figure caption for details.
Notice that the Dirac results agree with the TB ones, especially for the
lower-energy states.
In the case of an isolated ring, there are some discrepancies between the
two models for the hole energy levels.
This is due to the fact that in the TBM, applying IM boundary at the edge
boundary which is now not a perfect circle, breaks the e-h symmetry.
This symmetry, however, is preserved in the Dirac model as a result of its
perfect circular geometry.
Moreover, in the TBM results of the hybrid ring
[solid blue curves in Fig.\ \ref{fig6}(c)],
we see a bunch of energy levels decreasing in energy and approaching the zero energy.
They are known as the quantum \textit{Hall} edge states \cite{daCosta2016}
and are confined at the edges of the graphene flakes and are a result
of the finite size of the flake.
These Hall edge states are absent in the Dirac model for which the first layer 
is considered as an infinite graphene sheet.
In general, we expect good agreement between the Dirac and TB models for the
low-energy states.
For the higher energy levels and high magnetic field,
discrepancy between the Dirac and TB results becomes more significant.
The reason is that the linear spectrum invoked in the Dirac equation
is no longer valid.

We also notice that the energy spectra in both cases,
exhibit periodic oscillations as the magnetic flux varies.
This is a direct consequence of the AB effect \cite{Aharonov1959}
which is well known and
has been investigated for QRs (metallic, semiconductor as well as
graphene) \cite{Webb1985,Chakraborty1994,Russo2008,Wurm2010,Schelter2010}.
Similar to the AB
effect in semiconductor QRs, the energy oscillations
manifest itself in AB oscillations in the persistent current
$ j(\Phi) = - \frac{\p}{\p \Phi} E $.

\begin{figure}
	\centering
	\includegraphics[width=8.2cm]{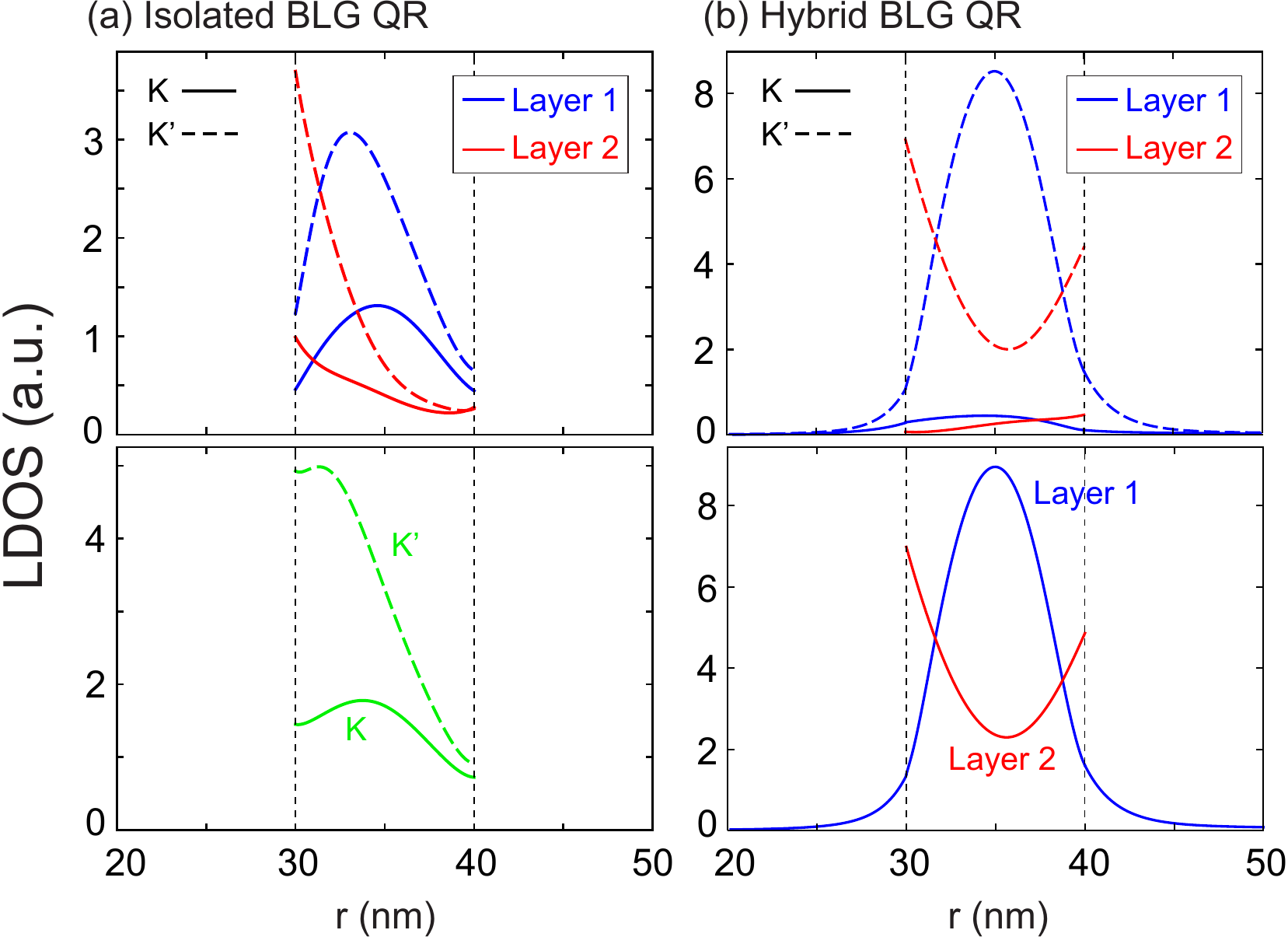}
	\caption[qw]{Valley- and layer-resolved local density of states
		[LDOS (arbitrary units)] of
		(a) an isolated BLG QR and (b) a hybrid BLG QR with the inner and outer
		radii of $ R_1 = 30 $ nm and $ R_2 = 40 $ nm, respectively.
		The magnetic field is $ B = 10 $ T.
		Layer 1 (2) is represented by the blue (red) curves.
		The solid (dashed) curves correspond to the $ K $ ($ K' $) valley.
		For the isolated BLG QR, the LDOS is computed in the energy
		range of $ [0, 0.2] $ eV, while for the hybrid one, the states in
		the energy window of $ [0, 0.11] $ eV
		[states between the lowest MLG LLs, $ n=0 $
		and $ n=1 $ in Fig.\ \ref{fig3}(c)] are sampled.
	}
	\label{fig7}
\end{figure}

Finally, we consider the valley- and layer-resolved LDOS for the studied
structures which can be probed by quantum capacitance
measurements and by scanning tunneling microscopy.
We consider the LDOS in the energy window of
$ \Omega \equiv [E_1, E_2] $ which can be obtained using
\begin{equation}
\rho_{\ell}^\tau (\mathbf{r}) = \sum_{E_i \in\, \Omega } \delta(E-E_i)
|\psi_\ell^{E_i}(\textbf{r})|^2,\
\label{eqrho}
\end{equation}
where $\psi_\ell^{E_i}(\textbf{r})=
[\phi_{A\ell}^{E_i}(\textbf{r}),
\phi_{B\ell}^{E_i}(\textbf{r})]^T$
denotes the quantum state of the two layers ($\ell=1,2$) with
energy $E_i$.
The components $\phi_{A\ell}(\textbf{r})$ and
$\phi_{B\ell}(\textbf{r})$
correspond, respectively, to the different sublattices $A\ell$ and
$B\ell$ in each layer at the given valley ($ \tau = K,K' $).

In Fig.\ \ref{fig7}, we show the valley- and layer-resolved LDOS
of the lowest-electron-energy states for both ring structures at $ B=10 $ T.
In the case of isolated BLG QR [Fig.\ \ref{fig7}(a)], for which the LDOS is
computed in the energy range of $ [0, 0.2] $ eV, one can see that the LDOS
of layer
1 (blue curves) is mostly centered inside the ring whereas in layer 2 the LDOS
has its maximum value at the inner side of the ring.
This behavior happens for both valleys, $ K $ and $ K' $
(solid and dashed curves, respectively).
Besides, in each layer, the maximum contribution in the LDOS belongs to the
$ K' $ valley.
The lower panel of Fig.\ \ref{fig7}(a) shows the LDOS for each valley separately.
We see that the LDOS contribution of the $ K' $ valley at the inner boundary
of the ring is dominant which decreases with radial distance and at the
outer side of the ring, contributions of both valleys in the LDOS are
almost equal.
Shown in panel (b) of Fig.\ \ref{fig7} is the LDOS of the hybrid BLG QR.
In this case, we compute the LDOS over the electron-energy states
located between the lowest MLG LLs, $ n=0 $ and $ n=1 $.
Here too, the contribution of the $ K' $ valley in the LDOS is dominant
in both layers.
While the LDOS of the lower layer (blue curves) is centered in the middle
of the ring, in the upper layer (red curves) the LDOS form rings with
maximums at the inner and outer boundaries of the ring.
Thus the difference between the
valley contributions as well as from the different layers in the LDOS at 
a given magnetic field can open up novel ways for applications of such 
graphene-based nanostructures in valleytronics.


\section{Summary and concluding remarks} \label{con}
In summary, based on continuum (Dirac-Weyl equation) and tight-binding models,
we studied the electronic properties of BLG QR, defined by IM potential, in two
different configurations: an isolated BLG QR and a hybrid one where
a MLG ring is put on top of an infinite MLG sheet.
By using the Dirac approximation and applying the IM boundary condition,
we first obtained analytical results for the energy levels and
corresponding wave spinors for both structures as function of a
perpendicular magnetic field.

In contrast to the previously investigated zero-width BLG QR, here,
the isolated BLG QR features a sizeable and magnetically tunable band gap
that decreases as the magnetic field strength is increased.
Our analytical findings are in excellent agreement with the tight-binding 
transport simulations.
Further, the theoretical results show the intervalley symmetry 
$E^K_e(m) = -E^{K'}_h (m)$ between the electron and hole states
for the energy levels of the isolated BLG QR, where $ m $ is the angular momentum
quantum number and $ K,K' $ refer to the two Dirac valleys.

The results for hybrid BLG QRs showed that the presence of an interface 
boundary in a hybrid
BLG QR modifies drastically the energy levels as compared to that of an isolated
BLG QR and its energy levels interplay between the MLG dot, isolated BLG QR, 
and MLG Landau energy levels as magnetic field varies.
No symmetry between the energy levels is found in this case.


Further, the energy spectrum of both structures exhibits Aharonov-Bohm
oscillations as the magnetic field varies.
We also explicitly confirmed the validity of our results by
simulating the QRs by a staggered site-dependent potential
in the TBM.
We found good agreement between our analytical results
obtained in the continuum approximation and those calculated within
the TBM.
Finally, we analyzed the spatial dependence of the valley- and
layer-resolved LDOS for the proposed BLG QRs.
Our findings are relevant for valleytronics applications.
Our results can be realized experimentally using
different techniques such as magnetotransport measurements similar to those 
reported in Refs.\ \cite{Jessen2019,Cabosart2017} for a MLG ring or 
using scanning probe techniques such as Scanning Tunneling Spectroscopy 
and  Scanning Gate Microscopy \cite{Cabosart2017} to probe the LDOS.

Although many-body effects, such as those coming from electron-electron and
electron-phonon interactions, may appear in some experiments in certain regimes 
and initial conditions, thus affecting the confinement properties in 
BLG-based nanostructures \cite{Goossens2012,Allen2012,Knothe2020}, 
the analytical
solution proposed here allows us to have physical insights into the basic mechanisms behind the results, which is of fundamental importance for a theoretical
understanding of some electronic properties in BLG nanostructures. 
Moreover, recent experimental measurements of quantum confined states in BLG QDs 
by using scanning tunneling microscope 
\cite{Ge2020,Kaladzhyan2021F,Kaladzhyan2021S,Ge2021,Joucken2021prl,Joucken2021nl}
have been confirmed by single-particle tight-binding calculations, even in the presence 
of charge defects, impurities, dopants and adatoms \cite{Kaladzhyan2021S,Joucken2021prl,Joucken2021nl}.

\section*{Acknowledgements}
We gratefully acknowledge discussions with I.\ Snyman.
This work was supported by the Institute
for Basic Science in Korea (No.\ IBS-R024-D1).
D.R.C is supported by CNPq grant numbers 310019/2018-4 and 437067/2018-1.

\end{document}